\documentclass[twocolumn,showpacs,preprintnumbers, prc,superscriptaddress]{revtex4}
\usepackage{epsfig}
\usepackage{graphicx}
\usepackage{amsmath,amssymb,amsfonts}
\usepackage{array}
\usepackage{url}
\usepackage{hyperref}
\usepackage{lineno}
\usepackage{xspace}
\usepackage[usenames,dvipsnames]{color}
\newcommand{\sqsn}{\mbox{$\sqrt{s_{_{NN}}}$}\xspace}
\newcommand{\bef}{\begin{figure}}
\newcommand{\eef}{\end{figure}}
\newcommand{\bc}{\begin{center}}
\newcommand{\ec}{\end{center}}

\newcommand{\be}{\begin{equation}}
\newcommand{\ee}{\end{equation}}
\newcommand{\bea}{\begin{eqnarray}}
\newcommand{\eea}{\end{eqnarray}}

\begin{document}
\title{Effect of correlations on cumulants in heavy-ion collisions}
\author{D.~K.~Mishra}
\email {dkmishra@rcf.rhic.bnl.gov}
\affiliation{Nuclear Physics Division, Bhabha Atomic Research Center, Mumbai 
400085, India}
\author{P.~Garg}
\email {prakhar@rcf.rhic.bnl.gov}
\affiliation{Discipline of Physics, School of Basic Science, Indian Institute of 
Technology, Indore 452020, India}
\author{P.~K.~Netrakanti}
\affiliation{Nuclear Physics Division, Bhabha Atomic Research Center, Mumbai 
400085, India}

\begin{abstract}
We study the effects of correlations on cumulants and their ratios of 
net-proton multiplicity distributions which have been measured for central 
(0\%--5\%) Au + Au collisions at the Relativistic Heavy Ion Collider (RHIC). 
This effect has been studied assuming individual proton and anti-proton 
distributions as a Poisson or Negative Binomial Distribution (NBD). In-spite of 
significantly correlated production due to baryon number, electric charge 
conservation and kinematical correlations of protons and anti-protons, the 
measured cumulants of the net-proton distribution follow the independent 
production model. In the present work we demonstrate how the introduction of the 
correlations will affect the cumulants and their ratios for the difference 
distributions. We have also demonstrated this study using the proton and 
anti-proton distributions obtained from the HIJING event generator.

\end{abstract}

\maketitle

\section{Introduction}
\label{intro} 
In recent years, the Beam Energy Scan (BES) program at Brookhaven National 
Laboratory's Relativistic Heavy-Ion Collider (RHIC) has drawn much attention to 
map the quantum chromodynamics (QCD) phase diagram in terms of temperature ($T$) 
and baryon chemical potential ($\mu_{B}$)~\cite{Stephanov:1998dy}. Lattice QCD 
calculations combined with other theoretical models suggest that there should be 
a critical point where the phase transition line of first order originating from 
high $\mu_{B}$ ends~\cite{Stephanov:2004wx,Fodor:2004nz,Stephanov:1999zu}. 
Experimentally the location of the critical point can be measured by scanning 
the $T-\mu_{B}$ phase diagram. One can scan the $T-\mu_B$ plane by varying the 
center of mass energies of the colliding ions. 

The moments of the multiplicity distribution of conserved quantities are 
related to the correlation length ($\xi$) of the system and hence can be used to 
look for signals of a phase transition and critical 
point~\cite{Ejiri:2005wq,Bazavov:2012vg}. The variance ($\sigma^2$) of these 
distributions is related to $\xi$ as $\sigma^2 \sim 
\xi^2$~\cite{Stephanov:1999zu}, the skewness ($S$) goes as $\xi^{4.5}$ and the 
kurtosis ($\kappa$) is related as $\xi^7$ 
~\cite{Stephanov:2008qz,Gavai:2010zn,Cheng:2008zh}. Also, these quantities have 
been used to extract the freeze-out parameters of the system. For example, 
higher moments of net-charge distributions are used to extract  $\mu_{B}$ and 
are found to be in good agreement with methods using particle 
ratios~\cite{Borsanyi:2014ewa,Alba:2014eba,Adare:2015aqk}. Both the current 
experiments at RHIC (STAR and PHENIX), have reported their  measurements of 
higher cumulants for net-charge~\cite{Adamczyk:2014fia,Adare:2015aqk} and 
net-proton~\cite{ Adamczyk:2013dal} multiplicity distributions at 
different collision energies. The net-proton results from the STAR measurements 
are reasonably described by assuming independent production of protons and 
anti-protons, indicating that there are no apparent correlations between the 
protons and anti-protons for the observable presented~\cite{Adamczyk:2013dal}. 
In the independent production (IP) model, the measured cumulants of protons and 
anti-protons are used to construct the cumulants of the net-proton 
distribution. If the individual proton and anti-proton distributions are assumed 
to be Poisson distributions, the resultant net-proton distribution will be a 
Skellam distribution~\cite{BraunMunzinger:2011ta}. Poisson distributions fall 
into the class of ``integer valued $Le\acute{v}y$ processes" for which the 
cumulants of the distribution $P(n^+-n^-)$ of the difference of samples from 
positive ($n^+$) and negative ($n^-$) distributions $P(n^+)$ and $P(n^-)$, with 
cumulants $C_n^+$ and $C_n^-$, respectively, are
\begin{equation}
\label{eq:levy}
C_n =  C^{+}_n +(-1)^n C^{-}_n
\end{equation}
so long as the distributions are not correlated~\cite{Johnson,Nielsen}. This 
result is the same as if the distributions $P(n^+)$ and $P(n^-)$ are 
statistically independent. A similar exercise has been carried out in 
Refs.~\cite{Tarnowsky:2012vu,Westfall:2014fwa} using heavy-ion event generators 
such as HIJING and UrQMD assuming both the individual distributions to be a 
Poisson or a negative binomial distribution (NBD). Further, the moments of the 
distributions are related to the cumulants as: mean ($M$) = $C_1$; $\sigma^2 = 
C_2 = \langle (\delta N)^2\rangle$; $S = C_3/C_2^{3/2} = \langle (\delta 
N)^3\rangle/\sigma^3$ and 
$\kappa = C_4/C_2^2 = \langle (\delta N)^4\rangle/\sigma^4 - 3$, where $N$ is
the multiplicity and $\delta N$ = $N-M$. Hence, the ratios of the cumulants are 
related to the moments as: $M/\sigma^2 = C_1/C_2$, $S\sigma = C_3/C_2$, 
$\kappa\sigma^2 = C_4/C_2$ and $S\sigma^3/M = C_3/C_1$. 

Recently, measured cumulants of net-charge distributions by the PHENIX 
experiment show that individual positively and negatively charged hadron 
multiplicity distributions can be described by NBD for energies from 
$\sqrt{s_{NN}}$ = 7.7 to 200 GeV in Au + Au collisions~\cite{Adare:2015aqk}. 
Since NBD  also lies in the class of ``integer valued $Le\acute{v}y$ processes", 
it also follows Eq.~\ref{eq:levy}. Hence, cumulants calculated from the 
event-by-event (e-by-e) net-charge distributions agree with the cumulants 
obtained from individual positive and negative multiplicity distributions using 
Eq.~\ref{eq:levy}.
 
In Refs.~\cite{Netrakanti:2014mta, Luo:2014tga}, the individual cumulants are 
shown for different particle production mechanisms and it is observed that 
within the STAR kinematical acceptance, models satisfy the independent 
production model with the e-by-e measured distributions. An ideal hadron 
resonance gas model~\cite{Karsch:2010ck,Garg:2013ata} in the grand canonical 
ensemble by construction treats the susceptibility of net-protons in a similar 
way as they are treated in the IP model as: $\chi^{(n)}_{p-\bar p} = 
\chi^{(n)}_{p} + (-1)^n \chi^{(n)}_{\bar p}$ where, $\chi^{n}_{p-\bar p}$ is the 
$n^{th}$ order susceptibility for net-protons, $\chi^{(n)}_{p}$ and 
$\chi^{(n)}_{\bar p}$ are the $n^{th}$ susceptibilities for protons and 
anti-protons, respectively. The STAR collaboration has reported that the product 
of moments of net-proton distributions are found to have values close to 
expectations based on independent proton and anti-proton 
production~\cite{Adamczyk:2013dal}. However, it has been puzzling since then 
that in-spite of significantly correlated production due to baryon number, 
electric charge conservation and kinematical correlations of proton and 
anti-protons, why the measured cumulants follow the independent production 
model. In the present work, we demonstrate the results of such cases by 
considering the Poisson and NBD distributions for the particle production. We 
compare the results of cumulants and their ratios by making an e-by-e 
distribution to those which are derived from Eq.~\ref{eq:levy} after introducing 
the correlations.

The paper is organized as follows. In the following section, we discuss the 
method used to include the correlation in this study. In 
Section~\ref{sec:results}, the results for the observable $C_{1} /C_{2}$, $C_{3} 
/C_{2}$, $C_{4} /C_{2}$ and $C_{3} /C_{1}$ as a function of the correlation 
coefficients are presented for Poisson and NBD distributions along with their 
cumulants. The correlation effect is also discussed using the HIJING event 
generator. Finally in Section~\ref{sec:summary}, we summarize our findings and 
discuss the implications of this work to the current experimental 
measurements in high energy heavy-ion collisions.

\section{Method}
\label{sec:method}
Suppose we have two independently produced distributions $Y_1$ and $Y_2$, from 
which one can construct the distribution of the difference $(Y_1-Y_2)$ on an 
e-by-e basis. One can introduce the correlation between individual distributions 
by taking a third independently produced distribution $Y_{12}$, such that 
$(Y_1+Y_{12})$ and $(Y_2+Y_{12})$ are new distributions which are correlated as 
they share a common distribution $Y_{12}$. The difference of these two 
distributions ($Y_1+Y_{12}$) and ($Y_2+Y_{12}$) will be same as $(Y_1-Y_2)$. 
Therefore, in spite of a correlation, it is possible that the difference 
distribution remains the same as if there is no correlation. Let us define two 
independently produced bivariate Poisson distributions $X_1$  and $X_2$ as the 
joint distribution of the random variables as is given in~\cite{Johnson}
\begin{equation}
\label{eq2}
X_1= Y_1+Y_{12} ~\mathrm{and}~ X_2= Y_2+Y_{12}
\end{equation}
where $Y_1$, $Y_2$ and $Y_{12}$ are mutually independent Poisson random 
variables with the means $\lambda_1$, $\lambda_2$ and $\lambda_{12}$ 
respectively. It can be easily shown that $X_1$ and $X_2$ have Poisson 
distributions with means $\lambda_1$+$\lambda_{12}$ and 
$\lambda_2$+$\lambda_{12}$ respectively. In case of Poisson distributions the 
correlation coefficient between the bivariate distributions, $\rho(X_1, X_2)$ 
is defined by :
\begin{equation}
\rho(X_1, X_2) = \frac{Var(Y_{12})}{\sqrt{Var(X_1)Var(X_2)}} 
\end{equation}
Since the variance and the mean are the same for a Poisson distribution 
therefore,
\begin{equation}
\label{eq:coeff}
\rho(X_1, X_2) = \frac{\lambda_{12}} {\sqrt{(\lambda_1+\lambda_{12})
(\lambda_2+\lambda_{12})}}
\end{equation}
This definition of a bivariate Poisson distribution automatically provides a 
method to generate correlated Poisson random variates such that $X_1$ and $X_2$ 
have Poisson distributions with the specified mean values and the correlation 
coefficient $(\rho)>0$. The correlation coefficient is used as a measures of 
the degree of linear dependence between the two random variables $X_1$ and 
$X_2$. In the present work, we generate three independent Poisson distributions 
$Y_1$, $Y_2$, and $Y_{12}$ on an e-by-e basis using a Monte-Carlo technique, 
with carefully chosen means $\lambda_1$, $\lambda_2$ and $\lambda_{12}$, and 
obtain ($X_1, X_2$) through the two operations shown in Eq.~\ref{eq2}.
\bef[ht!]
\bc
\includegraphics[width=0.5\textwidth,height=4 cm]{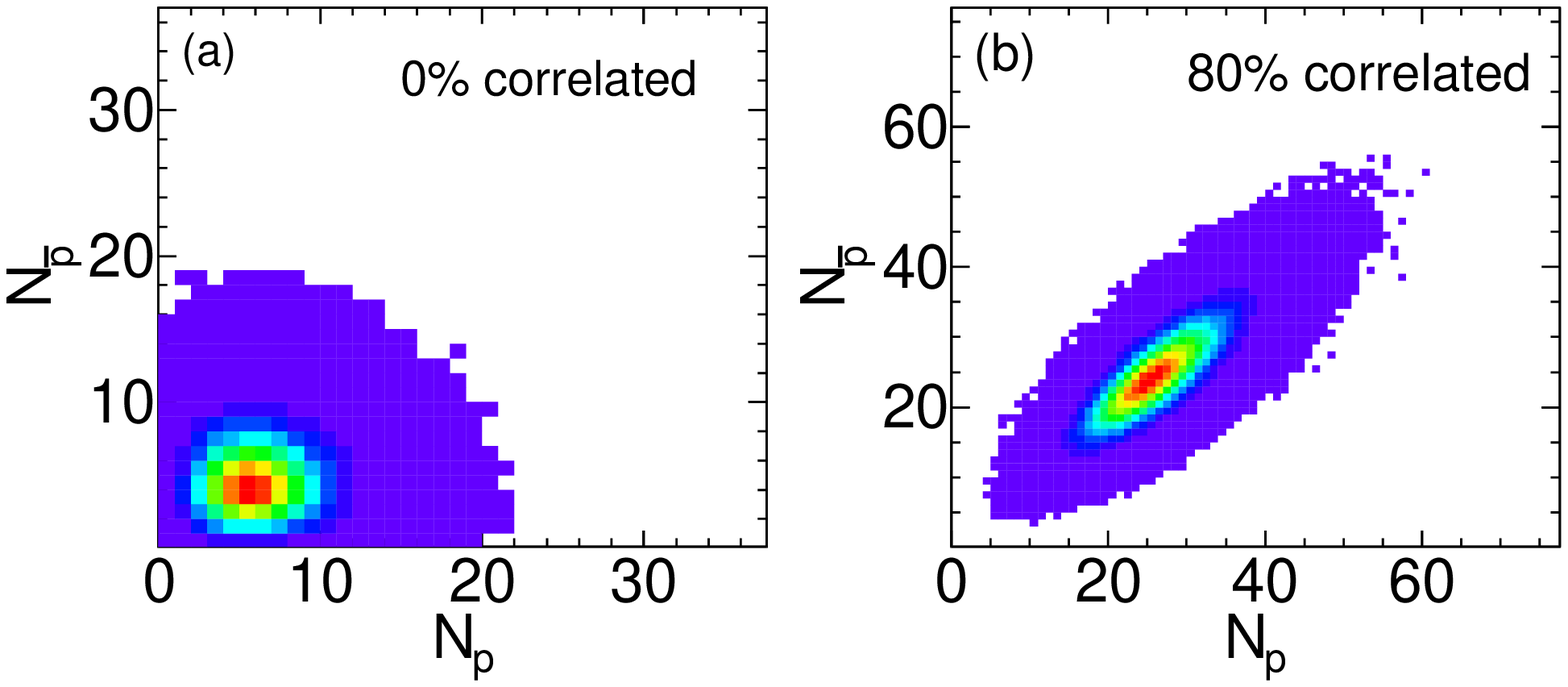}
\caption{Correlated $N_p$ and $N_{\bar p}$ distributions for a correlation 
coefficient ($\rho$) value of 0\% in panel (a) and 80\% in panel (b).}
\label{fig:posnegcorr}     
\ec
\eef
The mean values of the generated Poisson distributions $Y_1$ and $Y_2$ 
correspond to the mean of proton and anti-proton distributions for (0\%--5\%) 
centrality in Au + Au collisions at \sqsn = 19.6 and 200 GeV measured by the 
STAR experiment~\cite{Adamczyk:2013dal,stardata}. Afterwards, we chose a third 
distribution with a certain mean value ($\lambda_{12}$) so that we can control 
the correlation coefficient using Eq.~\ref{eq:coeff}. The two energies have 
been chosen in order to consider the highest RHIC energy along with a lower 
collision energy. We introduce the correlations by adding a third Poisson 
distribution to the individual proton and anti-proton distributions. Since the 
addition of two Poisson distributions is also a Poisson distribution, hence, 
both the distributions will remain Poisson after introducing the correlation. 
Figure~\ref{fig:posnegcorr} shows the typical correlation between the $N_p$ and 
$N_{\bar p}$ distributions for a correlation coefficient value of 0\% and 80\%. 
If both the distributions ($N_p$ and $N_{\bar p}$) will have the same mean and 
are uncorrelated ($\rho=0$), then the correlated distribution will have uniform 
circular distribution. The mean multiplicities used for $N_p$ are 5.664 $\pm$ 
0.0006 and 11.375 $\pm$ 0.003   for \sqsn =  200 and 19.6 GeV, and the mean 
multiplicities for $N_{\bar p}$ distributions are 4.116 $\pm$ 0.0005 and 1.15 
$\pm$ 0.001, respectively as given in Refs.~\cite{Adamczyk:2013dal,stardata}. 
We have repeated the same exercise assuming the individual $N_p$ and $N_{\bar 
p}$ distributions are given by NBDs. The negative binomial distribution function 
of an integer $n$ can be defined as:
\begin{equation}
 P(n) = \frac{\Gamma(n+k)}{\Gamma(n+1)\Gamma(k)}\frac{(\langle 
n\rangle/k)^n}{(1+\langle n\rangle/k)^{n+k}}
\end{equation}
where $\langle n\rangle$ is the mean number of particles and $k$ is an 
additional parameter. In the limiting case of $k\rightarrow\infty$, the NBD 
reduces to a Poisson distribution. The sum of two NBDs is also a negative 
binomial distribution. Figure~\ref{fig:meanvsrho} shows the mean values of the 
correlated (third) distribution which has been added to the individual $N_p$ and 
$N_{\bar p}$ distributions as a function of the correlation coefficient for 
two different energies \sqsn = 19.6 and 200 GeV. As we increase the mean value 
of the mixing distribution the correlation coefficient of the distributions 
also increase. However, there is very little energy dependence in 
$\lambda_{12}$ as a function of the correlation coefficient.

\bef[ht!]
\bc
\includegraphics[width=0.45\textwidth]{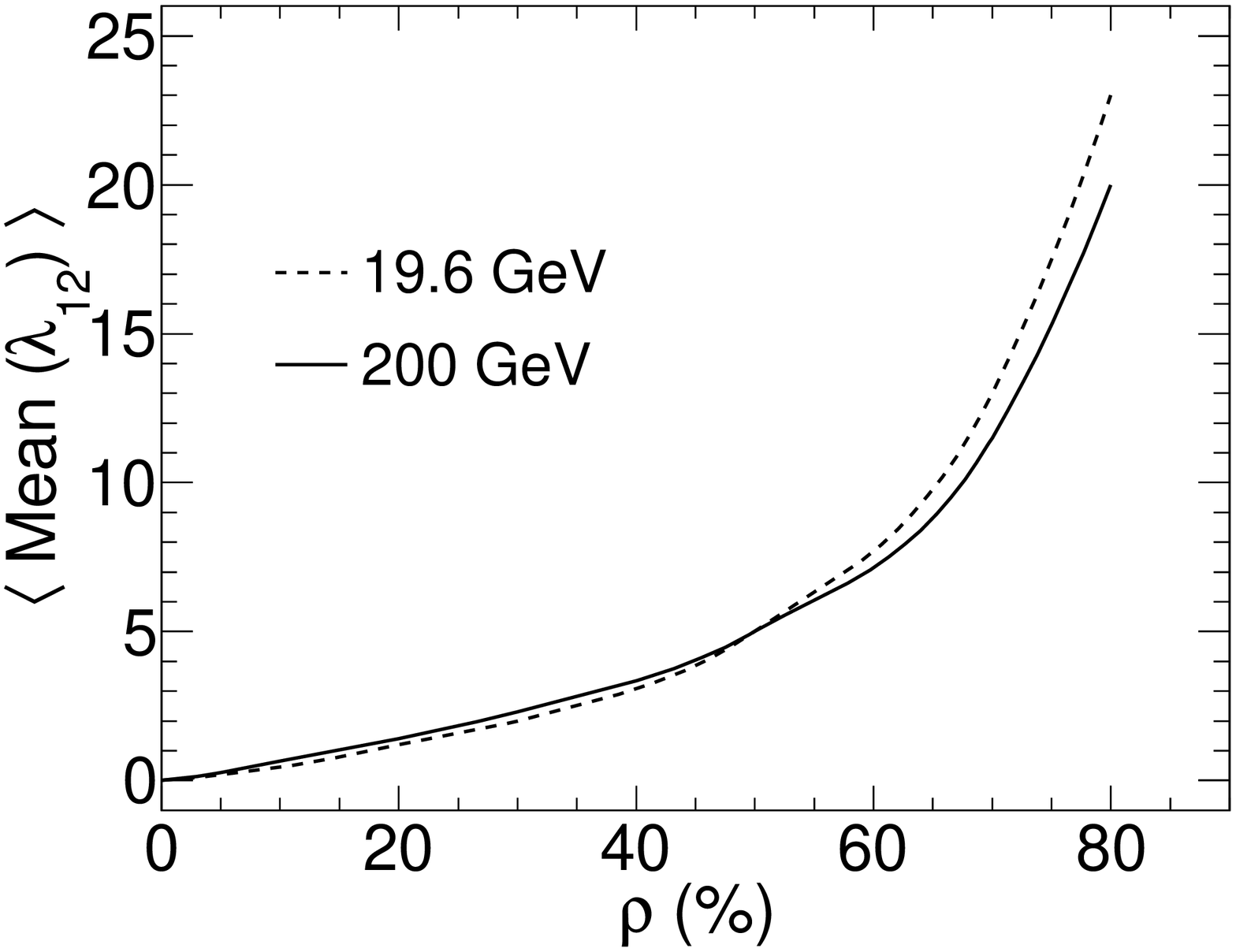}
\caption{The mean ($\lambda_{12}$) of the mixing distribution as a function of 
the correlation coefficient ($\rho$) for \sqsn = 19.6 and 200 GeV.}
\label{fig:meanvsrho}   
\ec
\eef
\section{Results and discussion}
\label{sec:results}
Experimentally, the measured higher moments of conserved quantities such as 
net-proton, net-charge and net-kaon are compared with baseline values, which are 
calculated by assuming the particle distributions as Poisson or negative 
binomial distributions. The Poisson statistics is a limiting case of NBD, where 
both the mean and variance of the distribution are the same. In the case of NBD 
the variance is larger than the mean of the distribution. In the following we 
will demonstrate the correlation effect on the higher moments and their ratios 
of net-multiplicities assuming individual positive and negative distributions 
as Poisson or NBD.

\subsection{Poisson distribution}
A statistically random expectation of the observable is described by Poisson 
distribution. The individual proton and anti-proton distributions are 
independently generated by using the measured mean values as given in 
Refs.~\cite{Adamczyk:2013dal, stardata}. Both the distributions are independent 
if the correlation coefficient is zero. We generate a third Poisson 
distribution by taking the different mean values which correspond to different 
$\rho$ as shown in Fig.~\ref{fig:meanvsrho}. Event-by-event we add the third 
distribution ($N_{mix}$) to the independently generated $N_p$ and $N_{\bar p}$ 
distributions. One can construct a corresponding net-distribution ($N_{diff}$) 
by taking the difference of the correlated ($N_p + N_{mix}$) and ($N_{\bar p} + 
N_{mix}$) distributions. The $N_{diff}$ distribution will be a Skellam 
distribution. In the present study, the cumulants of the net-distribution are 
calculated in two different ways. In first case cumulants of the 
net-distribution are calculated from the Skellam $N_{diff} (= N_p - N_{\bar p})$ 
distribution which is built by taking the correlated $N_p$ and $N_{\bar p}$ 
distributions on an e-by-e basis. In the second case the cumulants are 
calculated assuming independent production of particles as given in 
Eq.~\ref{eq:levy}.
\bef[ht!]
\bc
\includegraphics[width=0.5\textwidth]{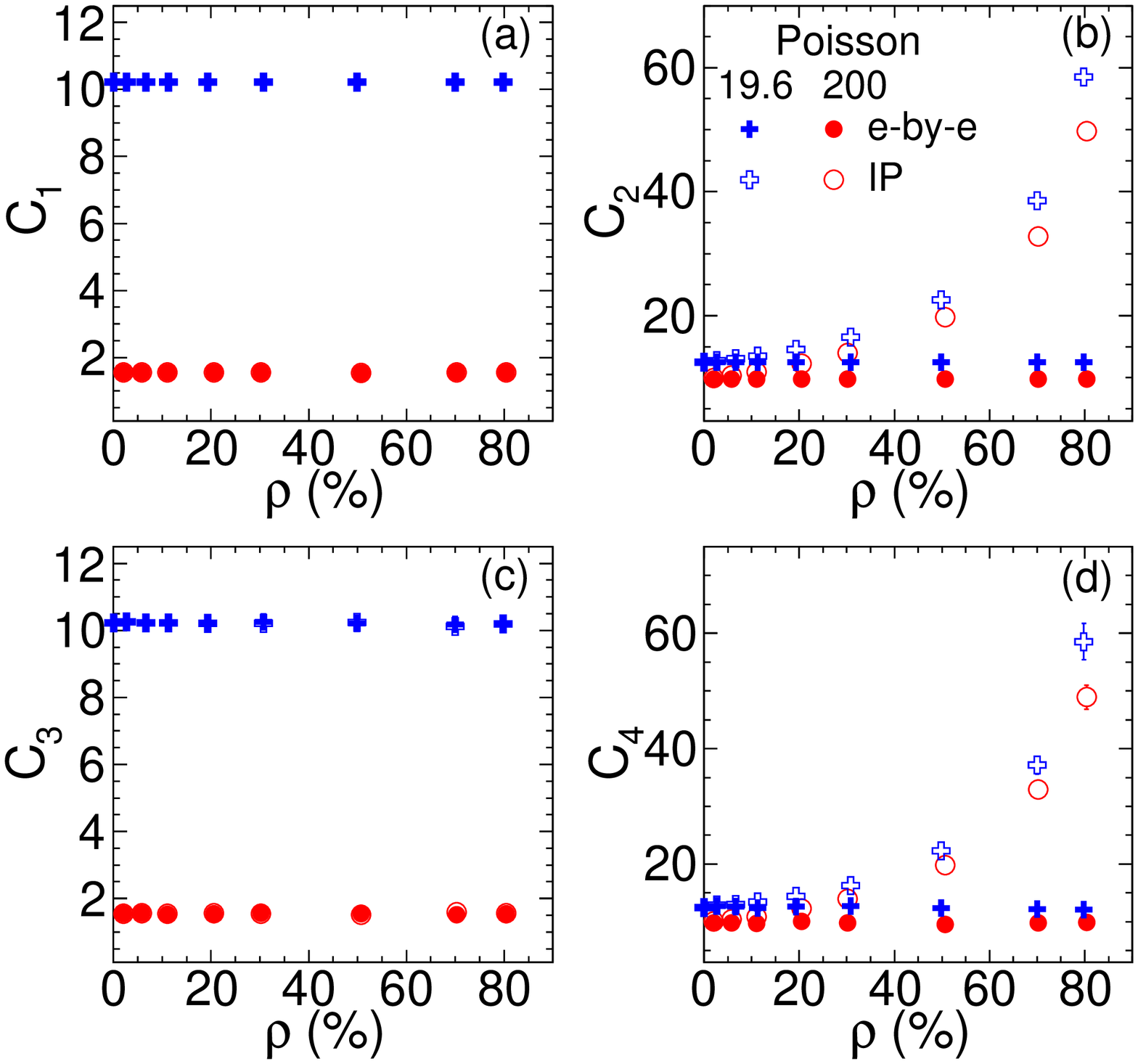}
\caption{Cumulants of net-proton distributions obtained on an 
e-by-e basis and by assuming independent production of particles are
shown as a function of the correlation coefficient ($\rho$) for \sqsn = 19.6 
and 200 GeV. The errors on the cumulants are statistical only and smaller than 
the symbol.}
\label{fig:cumu_pois}
\ec
\eef
Figure~\ref{fig:cumu_pois} shows the comparison of the cumulants calculated from
the e-by-e $N_{diff}$ distributions and by assuming independent 
production as a function of the correlation coefficient for \sqsn = 19.6 and 
200 GeV. As the measured mean ($C_1$) of proton and anti-protons are used to 
construct the Poisson distribution of $N_p$ and $N_{\bar p}$, the $C_1$ of the 
difference distribution calculated from the IP model and from e-by-e measured 
$N_{diff}$ distribution agrees. The $C_2$ and $C_4$ values obtained from the 
e-by-e $N_{diff}$ distributions are independent of $\rho$ as $N_{diff}$ 
is generated by taking the difference of $(N_p + N_{mix})$ and $(N_{\bar p} + 
N_{mix})$. However, there is strong dependence of $C_2$ and $C_4$ values as a 
function of $\rho$ when obtained using the IP model. As one increase the 
correlation 
coefficient, the $C_2$ and $C_4$ values deviate from the uncorrelated values. It 
is observed that, if two distributions are correlated more than $\sim$ 20\%, the 
$C_2$ and $C_4$ do not follow the ``integer valued $Le\acute{v}y$ processes". 
The $C_3$ values are independent of the correlation coefficient similar to the 
$C_1$ of the net-proton distribution. Similar behavior is observed for both 
\sqsn = 19.6 and 200 GeV. The statistical uncertainties for the various 
cumulants and their ratios are calculated using Delta theorem 
method~\cite{Luo:2011tp}. As noted in previous section, the uncertainties on the 
mean values of the protons and anti-protons are very small which gives 
negligible effect (less than 0.3\%) on the higher cumulants and their ratios. 
More discussion on the statistical uncertainties on higher cumulants also can be 
found in~\cite{Tarnowsky:2012vu}.
\bef[ht!]
\bc
\includegraphics[width=0.5\textwidth]{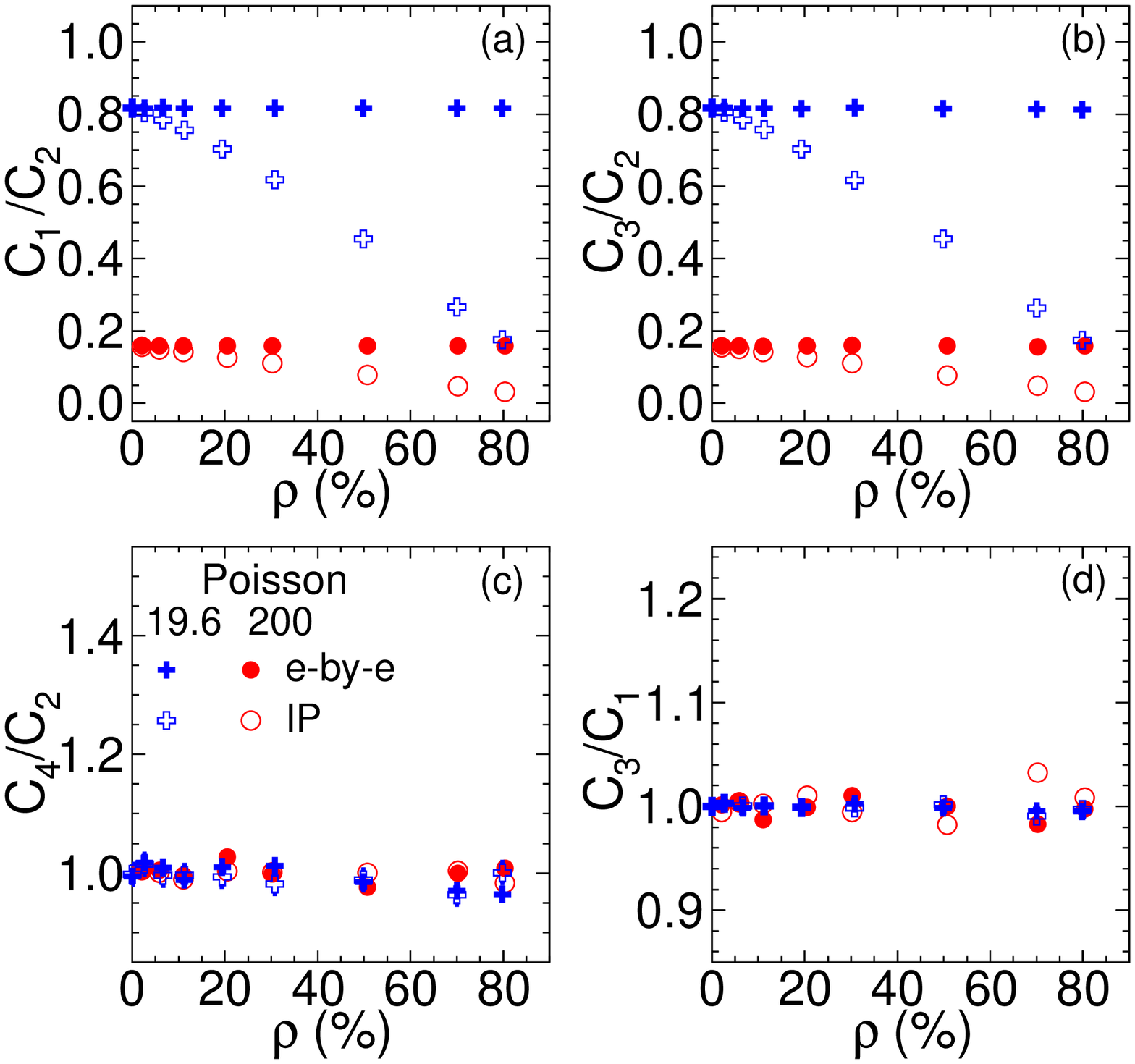}
\caption{Ratios of cumulants of net-proton distributions obtained on an 
e-by-e basis and by assuming the independent production of particles are shown 
as a function of the correlation coefficient ($\rho$) for \sqsn = 19.6 and 200 
GeV. The errors on the cumulants are statistical only and smaller than the 
symbol.}
\label{fig:cumu_pois_ratios} 
\ec
\eef
Figure.~\ref{fig:cumu_pois_ratios} shows the ratios of cumulants as a function 
the correlation coefficient for \sqsn = 19.6 and 200 GeV. The cumulant ratios 
obtained from the e-by-e measured $N_{diff}$ distributions are 
independent of $\rho$ as $N_{mix}$ distribution gets canceled-out while 
constructing the net-distribution. In case of independent production model, the 
correlation added to the individual $N_p$ and $N_{\bar p}$ distributions 
preserve. The $C_1/C_2$ and $C_3/C_2$ ratios show strong dependence of $\rho$ 
calculated using Eq.~\ref{eq:levy}. However, the $C_4/C_2$ and $C_3/C_1$ ratios 
are independent of $\rho$ in both the cases. Like individual cumulant case, the 
$C_1/C_2$ and $C_3/C_2$ ratios for very small $\rho$ starts deviating from the 
uncorrelated baseline ratios. Although there is larger correlation between the 
particles which are produced close to the critical end point (CEP) or phase 
transition, it will be difficult to observe in $C_4/C_2$ and $C_3/C_1$ ratios as 
these two ratios are independent for any degree of correlation. This indicates 
that, in heavy-ion collisions, even if the particles have small correlation, it 
can be seen in $C_1/C_2$ and $C_3/C_2$ ratios.
\bef[ht!]
\bc
\includegraphics[width=0.5\textwidth]{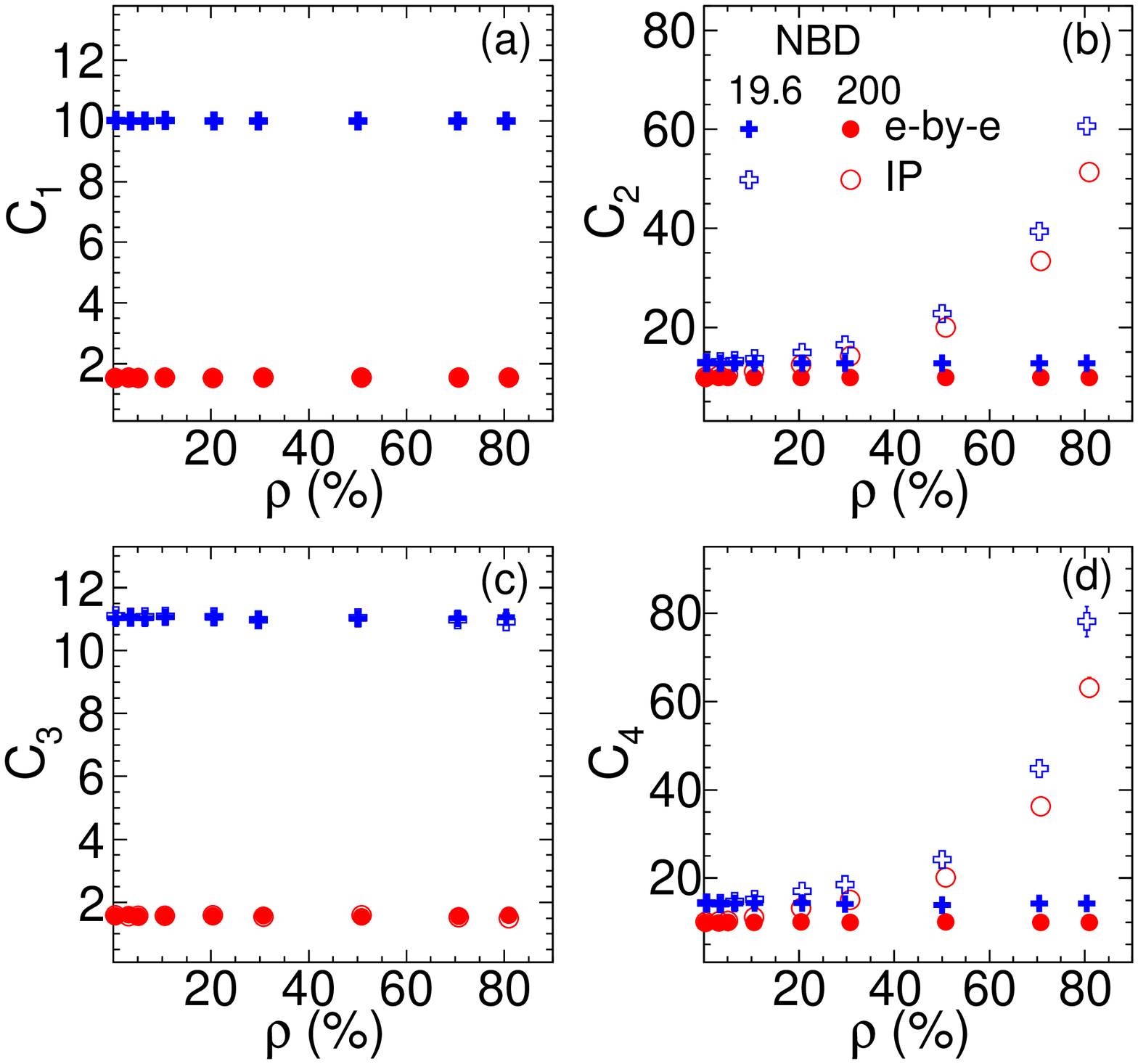}
\caption{Cumulants of net-proton distributions obtained on an e-by-e basis and 
by assuming independent production of particles are shown as a function of the 
correlation coefficient ($\rho$) for \sqsn = 19.6 and 200 GeV. The errors on the 
cumulants are statistical only and smaller than the symbol. The errors on the 
cumulants are statistical only and smaller than the symbol.}
\label{fig:cumu_nbd} 
\ec
\eef
\subsection{Negative Binomial Distributions}
The particle multiplicity distribution in elementary ($e^++e^-$ or $p+p$) as 
well as heavy-ion collisions are well described by negative binomial 
distribution~\cite{Alner:1985zc,Abbott:1995as,Becattini:1996,Adare:2008ns}. In 
the present study, the individual $N_p$ and $N_{\bar p}$  distributions are 
assumed to be negative binomial distributions, which are constructed by taking 
the measured $C_1$ and $C_2$ of the proton and anti-proton distributions at 
\sqsn = 19.6 and 200 GeV as given in Ref.~\cite{Adamczyk:2013dal,stardata}. It 
is assumed that the individual $N_p$ and $N_{\bar p}$ distributions are produced 
independently. As it is known that sum of two NBDs are also negative binomial 
distribution. A third NBD distribution ($N_{mix}$) has been added to the 
individual $N_p$ and $N_{\bar p}$ distributions on an e-by-e basis so that the 
resulting proton and anti-proton distributions will be correlated. The mean 
values of the correlated distribution correspond to different correlation 
coefficient as shown in Fig.~\ref{fig:meanvsrho}. Similar to Poisson 
distribution, the cumulants are calculated in two different ways as discussed 
before. 
\bef[ht!]
\bc
\includegraphics[width=0.5\textwidth]{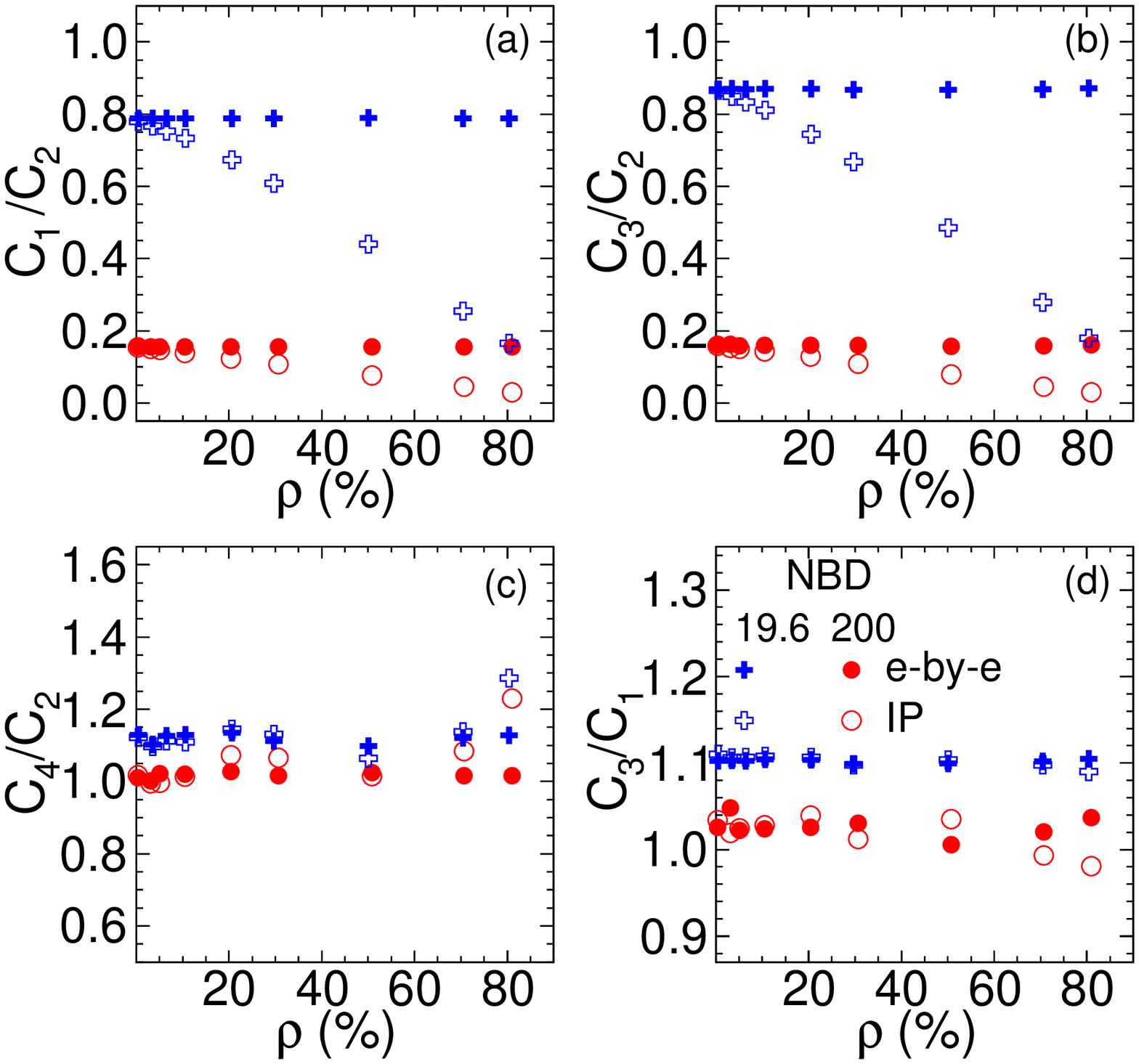}
\caption{Ratios of cumulants of net-proton distributions obtained on an 
e-by-e basis and by assuming independent production of particles are shown as a 
function of the correlation coefficient ($\rho$) for \sqsn = 19.6 and 200 GeV. 
The errors on the cumulants are statistical only and smaller than the symbol.}
\label{fig:cumu_nbd_ratios}
\ec
\eef
Figure~\ref{fig:cumu_nbd} shows the cumulants of the net-proton distributions 
calculated in both the methods as a function of the correlation coefficient 
for \sqsn = 19.6 and 200 GeV. Similar to the case of Poisson distribution, the 
$C_1$ and $C_3$ of net-proton distributions are obtained from both the methods 
are independent of $\rho$. Deviations from the uncorrelated baseline are 
observed for $C_2$ and $C_4$ of the net-proton distributions calculated in the 
IP model. As we increase the correlation coefficient the deviation of $C_2$ and 
$C_4$ increase from the baseline values. These two cumulants behave as 
uncorrelated until the correlation coefficient is less than $\sim$ 20\%. 
Figure~\ref{fig:cumu_nbd_ratios} shows the ratios of the cumulants as a function 
of the correlation coefficient for \sqsn = 19.6 and 200 GeV. The cumulant 
ratios calculated from the e-by-e $N_{diff}$ distributions are 
independent of $\rho$. Where as the $C_1/C_2$ and $C_3/C_2$ ratios of the 
net-proton distributions calculated from the individual $N_p$ and $N_{\bar p}$ 
distributions deviate from the uncorrelated baseline values. As one increases 
the correlation coefficient, the $C_1/C_2$ and $C_3/C_2$ ratios deviate from 
the uncorrelated values. The $C_4/C_2$ and $C_3/C_1$ ratios are found to be 
independent of the $\rho$ for both the cases.

The $C_4/C_2$ and $C_3/C_1$ ratios from the Poisson distribution and NBD are 
independent of $\rho$ which implies, although particles are strongly correlated 
in heavy-ion collisions, still the cumulant ratios of net-proton distribution 
can be explained by independent particle production model. On the other hand, 
$C_1/C_2$ and $C_3/C_2$ ratios calculated using the IP model are strongly 
dependent 
on the correlation coefficient. If the particles produced in heavy-ion 
collisions close to the CEP are highly correlated, that can be observed in the 
$C_1/C_2$ and $C_3/C_2$ ratios. However, in the present study we have simulated 
the correlation as an independent Poisson or NBD distribution.This correlation 
may not be the same as that from the QCD based arguments, about the 
sensitivities of the higher moments which are based upon the expected critical 
behavior of the correlation length~\cite{Cheng:2008zh,Gavai:2010zn}. From the 
above study, we show that $C_1/C_2$ and $C_3/C_2$ ratios are more sensitive to 
the correlation coefficient.
\bef[ht]
\bc
\includegraphics[width=0.5\textwidth]{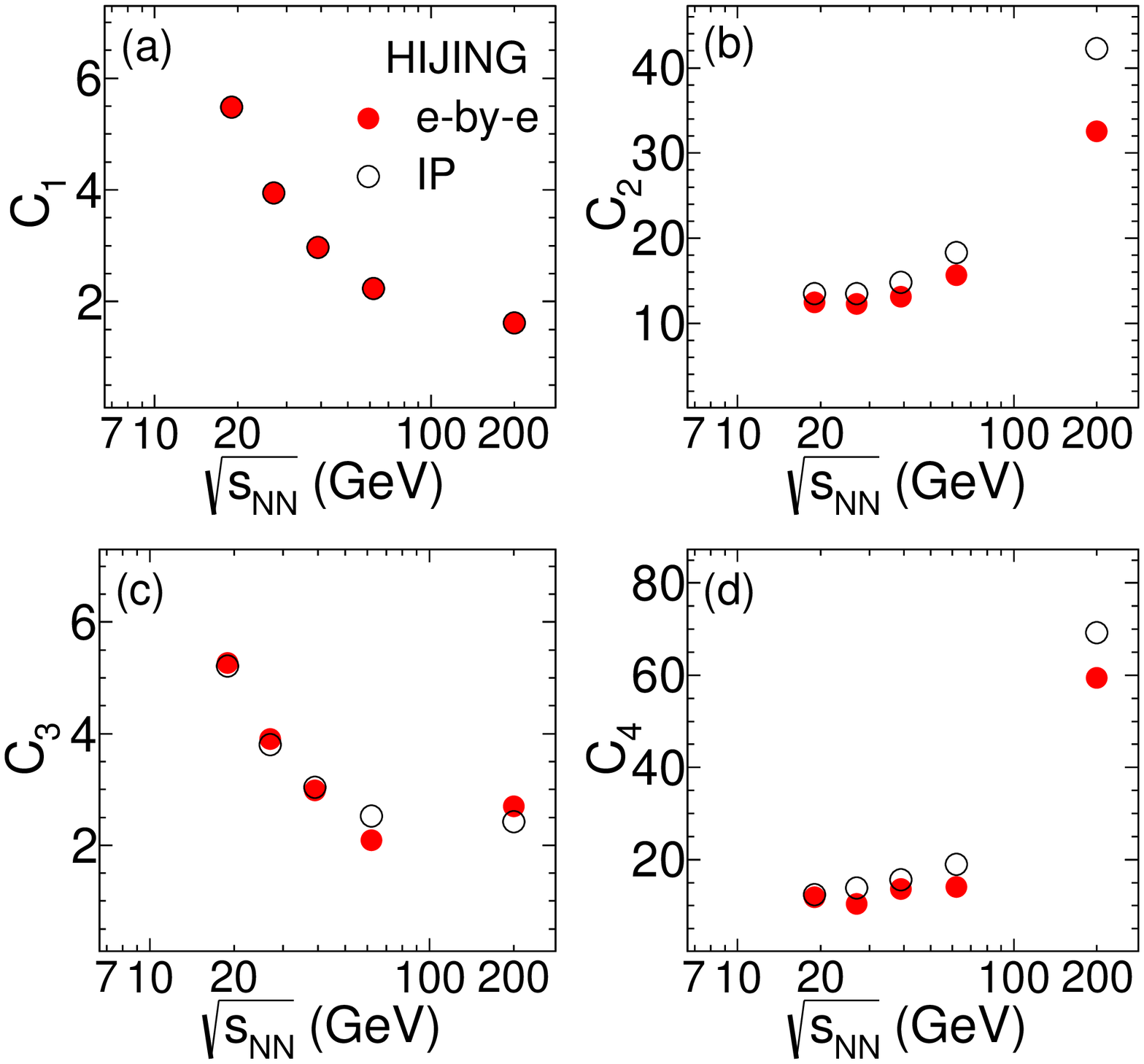}
\caption{Cumulants of net-proton multiplicity distributions obtained on an 
e-by-e basis and by assuming independent production of particles as a function 
of \sqsn for (0\%--5\%) centralities in Au + Au collisions from the HIJING. The 
errors on the cumulants are statistical only and smaller than the symbol.}
\label{fig:hijing_cumu}
\ec
\eef
\bef[h!]
\bc
\includegraphics[width=0.5\textwidth]{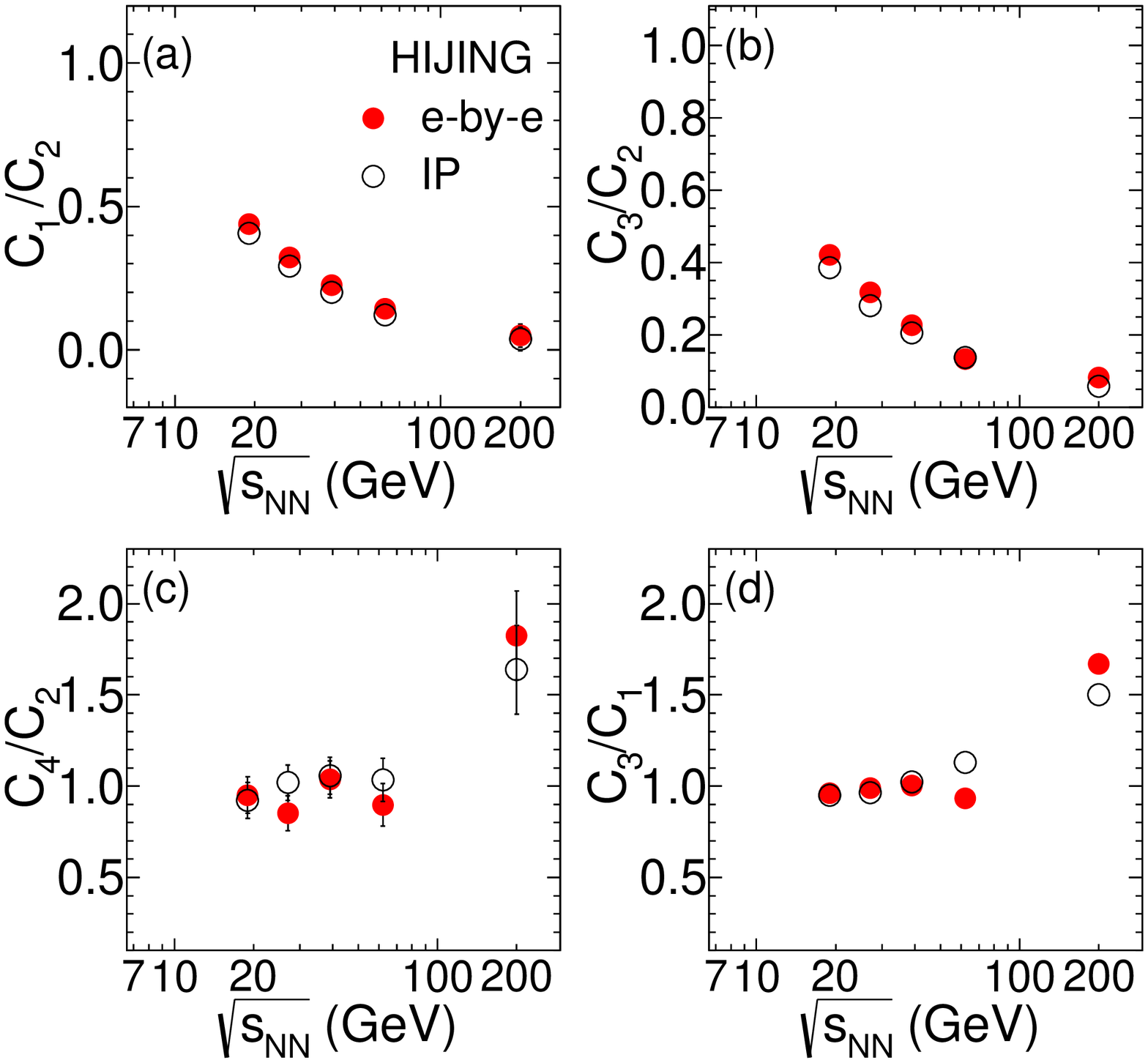}
\caption{Ratios of cumulants of net-proton multiplicity distributions obtained 
on an e-by-e basis and by assuming independent production of particles as a 
function of \sqsn for (0\%--5\%) centralities in Au + Au collisions from the 
HIJING.}
\label{fig:hijing_ratios} 
\ec
\eef
\subsection{Understanding the correlation effect with the HIJING model}
It is observed in Ref.~\cite{Adamczyk:2013dal} that the experimental data on 
net-proton cumulant ratios have been well explained by the independent 
production model. It can be argued that at lower collision energies the 
cumulants of net-proton distributions are mostly dominated by the cumulants of 
the corresponding proton distribution as the number of anti-proton production 
is very small. The $\bar p/p$ ratios are $\sim$ 0.01 and $~\sim$ 0.06 at \sqsn 
= 9.2 and 17.3 GeV respectively~\cite{Abelev:2009bw,Alt:2006dk}. However this 
argument will not hold for higher \sqsn as it is known from the measured 
$\bar{p}/p$ ratio $\sim$ 0.77 at mid-rapidity in Au + Au collisions at \sqsn = 
200 GeV~\cite{Abelev:2008ab} that there is significant correlation between 
proton and anti-proton production. Here we have discussed this aspect using the
Heavy Ion Jet interaction Generator (HIJING) model~\cite{Wang:1991hta}. It is a 
perturbative QCD model, which produces minijet partons that later are 
transformed into string fragments that then fragment into hadrons. 
Figure~\ref{fig:hijing_cumu} shows the cumulants of net-proton distributions 
calculated using both e-by-e measured $N_{diff}$ and individual cumulants of 
protons and anti-protons using Eq.~\ref{eq:levy} for the most central 
(0\%--5\%) Au + Au collisions at different \sqsn. The results shown for 
net-proton distributions are for the same acceptance as the experimental 
data~\cite{Adamczyk:2013dal}. The cumulants from both the methods are in 
agreement, although there is small difference for $C_2$ and $C_4$ values as 
observed in Poisson and NBD cases. At higher collision energies, the $C_2$ and 
$C_4$ values calculated using the IP model are slightly higher than the 
cumulants obtained from the e-by-e $N_{diff}$ distribution. This indicates, the 
presence of more correlations between the protons and anti-protons at \sqsn = 
200 GeV.
\bef[ht]
\bc
\includegraphics[width=0.4\textwidth]{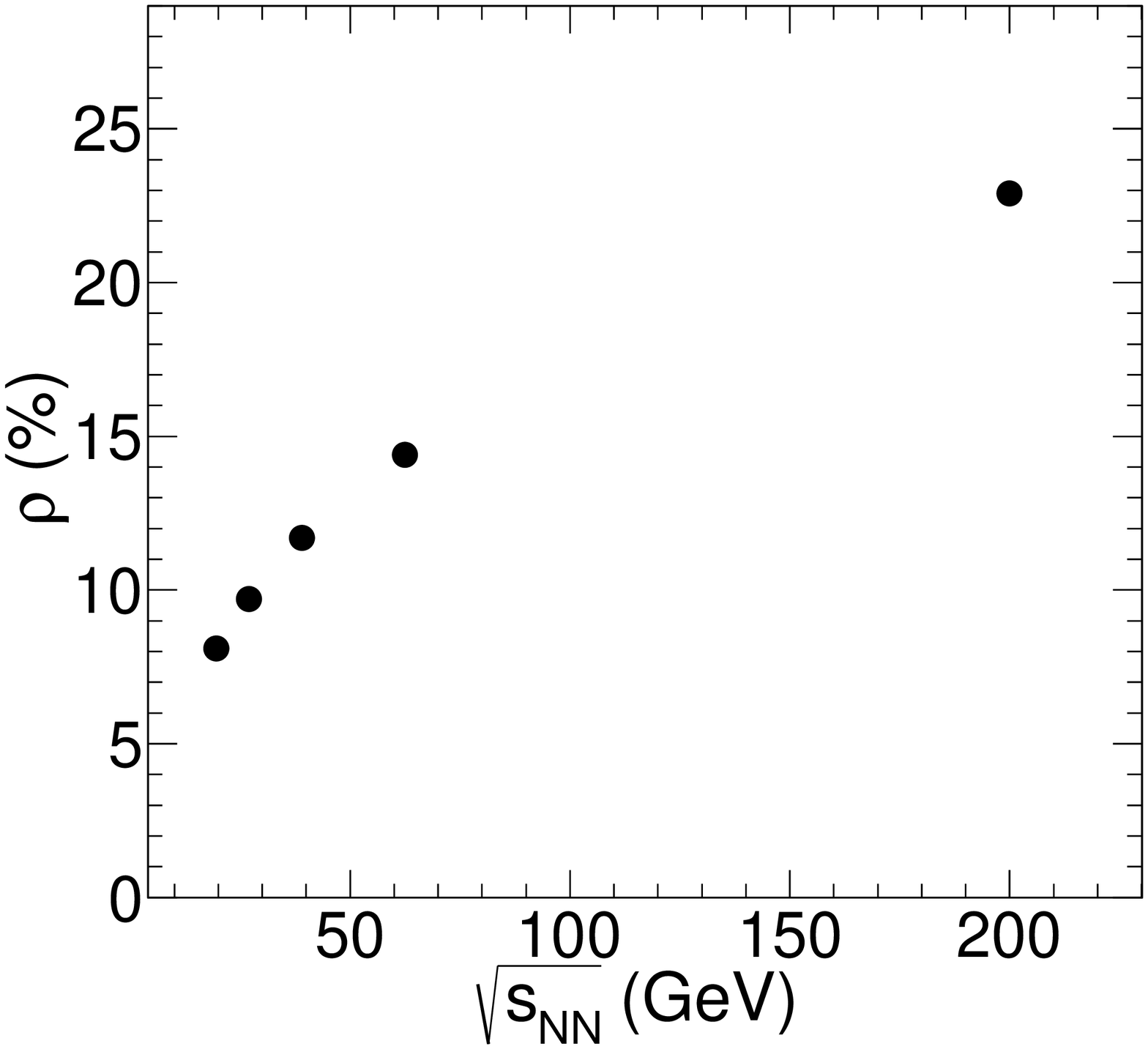}
\caption{Energy dependence of the correlation coefficient ($\rho$) calculated 
from $N_p$ and $N_{\bar p}$ distributions for (0\%--5\%) centrality in Au + Au 
collisions from the HIJING.}
\label{fig:enevsrho}
\ec
\eef
Figure~\ref{fig:hijing_ratios} shows the cumulant ratios as a function of 
\sqsn obtained from the above cumulants in Au + Au collisions for (0\%--5\%) 
centrality. The cumulant ratios calculated in both the methods agree very well. 
As we have seen in the cumulant ratios for Poisson and NBD cases, if the 
correlation exists between the proton and anti-proton then it should show in 
$C_1/C_2$ and $C_3/C_2$ ratios. We have observed in 
Fig.~\ref{fig:cumu_pois_ratios} and Fig.~\ref{fig:cumu_nbd_ratios} that the 
cumulant ratios calculated in both the methods agree even if the correlation 
coefficient $\sim$ 15\%. This implies, the correlation between protons and 
anti-protons in the HIJING events are within that order. Figure 
\ref{fig:enevsrho} shows the energy dependence of the degree of correlation that 
exists in proton and anti-proton production calculated from the HIJING model. 
As discussed previously in this section, even if the correlation coefficient 
is $\sim$ 20\%, the cumulants calculated in both the methods will agree, which 
has been observed in the case of the HIJING simulation. However, in all the 
studied cases (Poisson, NBD and HIJING), the $C_4/C_2$ and $C_3/C_1$ ratios 
calculated using both the methods agree. Hence, it is not surprising that the 
experimentally measured $C_3/C_2$ and $C_4/C_2$ ratios in 
Ref.~\cite{Adamczyk:2013dal} agree with the values calculated using the IP 
model. It may so happen that, after applying 
different kinematical cuts on the measurements within the experimental 
acceptance, the correlation coefficient values are reduced to less than 
$\sim20\%$. Experimental data that can be explained by the IP model does not 
rule-out the existence of CEP. We have demonstrated that even if particles are 
highly correlated the $C_4/C_2$ and $C_3/C_1$ ratios can be explained by the IP 
model. It is important to know how much correlation the protons and 
anti-protons should have so that one can claim to find the CEP. On the other 
hand, if experimentally measured particles have a correlation 
coefficient less than $\sim$ 20\%, the independent production model can explain 
the experimental cumulant ratios such as $C_1/C_2$ and $C_3/C_2$. 

\section{Summary}
\label{sec:summary}
In conclusion we have studied the effect of the correlations on the cumulants 
and their ratios assuming the experimentally measured proton and anti-proton 
distributions are described as a Poisson or NBD. The correlation is introduced 
in both the $N_p$ and $N_{\bar p}$ distributions. The cumulants and their ratios 
of net-proton distributions are calculated using e-by-e measured $N_{diff}$ 
distribution and from the independent production of $N_p$ and $N_{\bar p}$ 
distributions. We have demonstrated using Poisson and NBD distributions that, 
``integer valued $Le\acute{v}y$ processes" i.e $C_n$ = $C_n^+ + (-1)^n C_n^-$ 
for the net distribution is valid only if the correlation coefficient is less 
than $\sim$ 30\%--35\%. The $C_4/C_2$ and $C_3/C_1$ ratios are independent of 
the correlation coefficient, where as $C_1/C_2$ and $C_3/C_2$ ratios are more 
sensitive to the correlation coefficient. The $C_4/C_2$ and $C_3/C_1$ ratios 
can be explained by the IP model, which was also observed in 
Ref.~\cite{Adamczyk:2013dal}. The agreement between experimental data and the
independent production model is not a coincidence in which measurements have 
been carried out. We have discussed that, if the particles are highly 
correlated, one should look for $C_1/C_2$ and $C_3/C_2$ ratios as a function of 
collision energies, which will have larger deviations from the ratios obtained 
by uncorrelated baseline values. The cumulants and their ratios are calculated 
as a function of \sqsn using the HIJING event generator. In the HIJING model 
the cumulants calculated using both the methods agree very well. Hence, 
experimentally measured cumulants will follow the independent production model 
calculations if the correlation coefficient is less than $\sim$ 20\%. However, 
$C_4/C_2$ and $C_3/C_1$ values will follow the IP model for all the correlation 
coefficient values. The observation that the  experimental data can be 
explained by the independent production of particles does not rule-out the 
existence of the critical endpoint.


\end{document}